# Effects of Ru Doping on the Transport Behaviors and Superconducting Transition Temperature of $NdFeAsO_{0.89}F_{0.11}$

Sang Chul LEE,[1] Erika SATOMI,[1] Yoshiaki KOBAYASHI,[1,2] and Masatoshi SATO,[1,2]*

[1]*Department of Physics, Division of Material Science, Nagoya University, Furo-cho, Chikusa-ku, Nagoya 464-8602, Japan.*
[2]*JST, TRIP, Nagoya University, Furo-cho, Chikusa-ku, Nagoya 464-8602, Japan.*

The transport behavior and superconducting transition temperature $T_c$ of $NdFe_{1-y}Ru_yAsO_{0.89}F_{0.11}$ have been studied for various $y$ values. Because Ru impurities are isoelectronic to host Fe atoms, we basically expect that the number of electrons does not change with $y$, at least in the region of small $y$ values. The results indicate that the rate of $T_c$ suppression by Ru atoms is too small to be explained by the pair breaking effect of nonmagnetic impurities expected for the $S_\pm$ symmetry, confirming our previous results for Co doping.

**KEYWORDS: $NdFe_{1-y}M_yAsO_{1-x}F_x$ (M=Co, Ru), thermoelectric power, Hall coefficient, impurity effect**

For the newly found Fe pnictide superconductors[1)], various studies have been carried out to identify the symmetry of their superconducting order parameter Δ. In many of these studies, much effort has been made to find experimental evidence for the $S_\pm$ symmetry proposed theoretically at the early stage of the study.[2, 3)] For such a symmetry, reflecting the sign difference between the order parameters on disconnected Fermi surfaces around the Γ [= (0, 0)] and M [= (π, 0)] points in the reciprocal space, important features can be expected in several observable physical quantities: Neutron inelastic scattering measurements have been carried out to find the so-called "resonance peak" in the magnetic excitation spectra $\chi''(\boldsymbol{Q}, \omega)$[4-6)] expected in the superconducting phase around a point in the scattering vector($\boldsymbol{Q}$)-energy(ω) space,[7, 8)] and an observed peak has been discussed in relation to the "resonance peak". After the confirmation of the singlet state of Cooper pairs by Knight-shift measurements,[9, 10)] the temperature ($T$) dependence of the NMR longitudinal relaxation rate $1/T_1$ has been extensively discussed, and the absence of the coherence peak has been pointed out by many research groups.[11-15)] The $T$ dependence described by the relation $1/T_1 \propto T^n$ with $n$~3 has been reported below $T_c$ in almost the entire $T$ region studied for $LaFeAsO_{1-x}F_x$[11, 13, 14)] and $LaFeAsO_{1-\delta}$.[12)] These results have been discussed to be favorable for the $S_\pm$ symmetry.

The effects of impurity doping can provide information on the relative signs of the order parameters on Fermi surfaces around the Γ and M points, and on the basis of Co doping studies, we have emphasized that the observed rate of $T_c$ decrease due to the doping of nonmagnetic impurities is too small to be explained by the pair breaking effect expected for the $S_\pm$ symmetry of the order parameter.[9, 15-19)] The result seems to be consistent with those of studies carried out by neutron[20)] and α-particle[21)] irradiations. Onari and Kontani have pointed out from the theoretical side that the above data of doping effects cannot be explained by the pair breaking of nonmagnetic impurities.[22)] Regarding the magnetic excitation spectra $\chi''(\boldsymbol{Q}, \omega)$, it has been pointed[23)] out that a "peak" in $\chi''(\boldsymbol{Q}, \omega)$ is also expected for the $S$ symmetry of the order parameter, which has no sign difference between the Fermi surfaces around the Γ and M points, suggesting that we have to be careful in arguing whether the observed data really indicate the existence of the "resonance peak". On the $T$ dependence of the NMR relaxation rate $1/T_1$ of $LaFeAsO_{1-x}F_x$ and $LaFeAsO_{1-\delta}$, the observation[15, 19)] of the relation $1/T_1 \propto T^{5.5\text{-}6.0}$ indicates that the $1/T_1 \propto T^3$ stated above does not universally hold. We think that these two distinct $T$ dependences require more detailed consideration to extract information on the symmetry of Δ.

Various other studies have shown experimental evidence for the absence of nodes of the order parameter Δ,[24-27)] although they do not provide information if the sign change of the order parameters exists between two disconnected Fermi surfaces around the Γ and M points.

Here, to further extract information on the effects of nonmagnetic impurities, we have mainly studied the transport behavior and superconducting transition temperature $T_c$ of $NdFe_{1-y}Ru_yAsO_{0.89}F_{0.11}$ with changing $y$. In this case, because Ru impurities are isoelectronic to host Fe atoms, we basically expect that the number of electrons does not change with a change in $y$, at least in the region of small $y$ values. The results of the studies indicate that the rate of $T_c$ suppression by Ru doping is very small, confirming our previous results for Co doping that doped impurities do not seem to act as pair breaking centers, possibly excluding the $S_\pm$ symmetry of the order parameter. We also discuss the electronic state of the present system.

Polycrystalline samples of $NdFe_{1-y}Ru_yAsO_{0.89}F_{0.11}$ ($y$: nominal) were prepared from initial mixtures of Nd, $Nd_2O_3$, $NdF_3$, FeAs, and RuAs or Ru and As at nominal molar ratios. RuAs powders were obtained by annealing mixtures of Ru and As in an evacuated quartz ampoule at 850 °C. Samples of $LaFeAsO_{0.89-x}F_{0.11+x}$ were also prepared for the comparison of various physical properties with those of $NdFe_{1-y}Ru_yAsO_{0.89}F_{0.11}$. Details of the preparations and characterizations are given in our previous papers[9, 15-19)]. The X-ray powder patterns were taken with CuKα radiation at a step of 0.01° of the scattering angle 2θ. We found that the Ru doping was successful [The linear dependence of the lattice parameter $a$ on the Ru concentration $y$ shown in Fig. 1(a) guarantees that the actual $y$ value does not have a significant deviation from the nominal value, and therefore the deviation, even if it exists, does not affect the conclusion of the present paper.]. The superconducting diamagnetic moments were measured using a Quantum Design SQUID magnetometer with a magnetic field $H$ of 10 G under both zero-field-cooling (ZFC) and field-cooling (FC) conditions. In Fig. 1(b), only the ZFC data are shown. We note that the superconductivities of all the samples shown in the figure are not filamentary, because the observed diamagnetic moment is large. Electrical resistivity was measured by a four terminal method with an ac-resistance bridge. $T_c$ was determined as described previously[9, 16-19)] using the data in Fig. 1(b) and also those of the resistivities ρ shown in Figs. 2(a) and 2(b). The $T_c$ values obtained from these two kinds of data agree reasonably well.

*corresponding author (e43247a@nucc.cc.nagoya-u.ac.jp)

The Hall coefficient $R_H$ of the polycrystalline samples was measured with a stepwise increase in $T$ at a magnetic field $H$ of 7 T, where the sample plates were rotated around the axis perpendicular to the field, and the thermoelectric power $S$ was measured by the methods described in refs. 28 and 29.

Figures 2(a) and 2(b) show plots of the electrical resistivity against $T$ for the samples of $NdFe_{1-y}Ru_yAsO_{0.89}F_{0.11}$ with various $y$ values. One marked characteristic of the data is that the superconducting transition can be observed even for the samples with $y$= 0.5 and 0.6. Although the magnitudes of resistivity at 300 K do not exhibit a regular order with $y$, they have a rough tendency that they first increase with increasing $y$ from 0.0, and then decrease when $y$ exceeds 0.15. Figure 3 shows the residual resistivities of the samples with $y\leq$ 0.15 obtained by extrapolating the normal state resistivity data to $T$=0 from the $T$ region slightly above $T_c$. In the figure, the residual resistivities of the samples of $NdFe_{1-y}Co_yAsO_{0.89}F_{0.11}$ and $LaFeAsO_{0.89-x}F_{0.11+x}$ are also shown. (The data of $NdFe_{1-y}Co_yAsO_{0.89}F_{0.11}$ are from our previous paper.[19)] Detailed resistivity data for the latter system will be published in a separate paper.) From the figure, the rate of resistivity increase of $NdFe_{1-y}Ru_yAsO_{0.89}F_{0.11}$ with $y$ is almost equal to that of $NdFe_{1-y}Co_yAsO_{0.89}F_{0.11}$ in the limit of $y\to 0$, indicating that the scattering strengths of Ru and Co atoms within $NdFeAsO_{0.89}F_{0.11}$ are almost equal. With increasing $y$, the difference between the resistivities of Co- and Ru-doped systems becomes significant. This is because the difference between the electron numbers of these two kinds of systems becomes appreciable with increasing $y$. (Note that Co donates one electron into $NdFeAsO_{0.89}F_{011}$, while Ru does not.) In Fig. 3, we also show, for comparison, the residual resistivities of $LaFeAsO_{0.89-x}F_{0.11+x}$. For this system, the number of electrons doped into $LaFeAsO_{0.89}F_{0.11}$ increases with increasing $x$, and the scattering of conduction electrons by doped F atoms is expected to be weak, because F atoms are not in the conducting FeAs layers. From the data, we find that the resistivity increase induced by the F doping is not appreciable, and that the effect of Co or Ru doping on the residual resistivities of the present systems is apparent.

When $y$ exceeds ~0.15, residual resistivity has a tendency to decrease with increasing $y$. This cannot be understood within the framework of impurity effects. Instead, we have to consider the changes in various material parameters with $y$. Such effects of Ru substitution for Fe have been pointed out by McGuire *et al.*,[30)] for $PrFe_{1-y}Ru_yAsO$ for example, that is, the band width increases and Stoner factor decreases with increasing $y$.

The main panel of Fig. 4 shows the $T$ dependence of the thermoelectric powers $S$ of a series of $NdFe_{1-y}Ru_yAsO_{0.89}F_{0.11}$ samples with various $y$ values. In the smaller panel, the thermoelectric powers $S$ of a series of $NdFe_{1-y}Co_yAsO_{0.89}F_{0.11}$ samples already reported in our previous paper[19)] are also shown, for comparison. Although the absolute values of $S$ of both the Ru- and Co-doped systems exhibit decreasing tendencies with increasing $y$, we can see a characteristic difference between the changes of their $T$ dependences with $y$: For the Ru-doped system, the temperature $T_m$ at which the absolute $S$ has a maximum value remains almost constant in the $y$ region <0.15, while for Co-doped system, $T_m$ increases with $y$. We presume that this difference between the Ru and Co doped systems stems from the existence or nonexistence of a change in the number of electrons with $y$: As we pointed out previously,[18, 19)] the rather unusual $T$ dependence with a peak of $|S|$ reminiscent of those of high-$T_c$ Cu oxides can be considered to arise from the strong scattering of conduction electrons on electron Fermi surfaces around the (π, 0) point, by magnetic fluctuations of the holes in Fermi surfaces around the Γ point. For the Co- doped systems, along with the diminishing area of hole Fermi surfaces with increasing $y$, the magnetic fluctuation becomes weak and the unusual peak structure of the $|S|$-$T$ curves becomes less significant. For this reason, $S$ has a tendency to approach the ordinary $T$-linear behavior with the doping, resulting in an upward shift in $T_m$. For Ru doping, because hole Fermi surfaces do not become buried with increasing $y$,[30)] the peak structure remains with increasing $y$. The decrease in $|S|$ with increasing $y$ can be understood considering the weakening of the magnetic fluctuation of the system pointed out by McGuire *et al.*[30)] For $y\geq$0.5, we do not see any indication of strong magnetic fluctuation in the $T$ dependence of $S$. (Even for $y$=0.3, the effect of magnetism is rather weak.)

In the main panel of Fig. 5, the Hall coefficients of the series of samples of $NdFe_{1-y}Ru_yAsO_{0.89}F_{0.11}$ with various values of $y$ are shown against $T$. In contrast to the data of $NdFe_{1-y}Co_yAsO_{0.89}F_{0.11}$ shown in the smaller panel,[19)] strong $T$ dependences reminiscent of those of high-$T_c$ Cu Oxides can be observed for samples with $y$ as large as 0.15. This can be basically understood by considering that the number of electrons does not change with the Ru concentration $y$. For $y\geq$0.5, the $T$ dependence of $y$ is very weak. We can understand this behavior by considering the change in the strength of the magnetic fluctuations, as in the case of the thermoelectric power $S$.

In Fig. 6, we summarize the $T_c$ values of $NdFe_{1-y}Ru_yAsO_{0.89}F_{0.11}$ and $NdFe_{1-y}Co_yAsO_{0.89}F_{0.11}$ against $y$. In our previous paper,[19)] we estimated the rate of $T_c$ suppression by Co doping, adopting a pair breaking model for Co impurities, where it was found for material parameters deduced from the residual resistivities and Hall coefficients, that $T_c$ should vanish at the critical concentration $y_c$<0.01, which is much smaller than the observed values (~0.13; see Fig. 6). For the present Ru doping, the observed residual resistivities indicate that the scattering strength of Ru impurities is roughly equal to that of Co atoms in the region of small $y$ values. Therefore, the rate of $T_c$ suppression by Ru impurities expected in the present system should be nearly equal to the value for Co doping ($y_c$<0.01). Although this estimation may have certain errors, it is difficult to consider that nonmagnetic impurities act as pair breaking centers. In other words, the order parameter of the present superconducting system has no characteristic sign change of the $S_\pm$ symmetry. This argument is consistent with the theoretical result reported in ref. 22.

The present result obtained for Ru doping strengthens our previous results obtained in the study of Co doping. Here, we make a final comment on the difference in the rate of $T_c$ suppression between Co and Ru dopings. The superconductivity occurs in the presence of hole Fermi surfaces around the Γ point. Then, for Co doping, the superconductivity disappears, when the hole-Fermi-surfaces are buried by the electrons donated by Co impurities. This situation is different from the case of Ru doping, where the number of electrons remains constant with increasing $y$.[30] This difference may result in the difference in $y_c$ values between the Co- and Ru- doped systems.

In summary, we have presented the effects of Ru doping into $NdFeAsO_{0.89}F_{0.11}$ on the transport behavior and the superconducting transition temperature $T_c$, where the number of electrons is considered to remain constant with changing $y$, at least, in the region of small $y$ values. We have found that the rate of $T_c$ suppression is too small to be explained by the pair breaking effects of nonmagnetic impurities expected for superconductors with the $S_\pm$ symmetry. We expect that the present result will shed light on the superconducting symmetry of Fe pnictide systems.

Acknowledgments –The authors thank Prof. H. Kontani for fruitful discussion. The work is supported by Grants-in-Aid for Scientific Research from the Japan Society for the Promotion of Science (JSPS), and Technology and JST, TRIP.

## Figure captions

Fig. 1. (color online) (a) Lattice parameters $a$ and $c$ are plotted against $y$ of $NdFe_{1-y}Ru_yAsO_{0.89}F_{0.11}$. (b) The magnetic susceptibility of $NdFe_{1-y}Ru_yAsO_{0.89}F_{0.11}$ measured at an applied magnetic field $H$=10 G under the zero-field-cooling condition is shown against $T$ for samples with various $y$ values.

Fig. 2. (color online) (a), (b) The electrical resistivity of $NdFe_{1-y}Ru_yAsO_{0.89}F_{0.11}$ is shown against $T$ for samples with various $y$ values.

Fig. 3. (color online) The residual resistivities of $NdFe_{1-y}Ru_yAsO_{0.89}F_{0.11}$, $NdFe_{1-y}Co_yAsO_{0.89}F_{0.11}$, and $LaFeAsO_{0.89-x}F_{0.11+x}$ are shown against $y$ or $x$.

Fig. 4. (color online) The main panel shows the thermoelectric power of $NdFe_{1-y}Ru_yAsO_{0.89}F_{0.11}$ against $T$ for various $y$ values, and the smaller panel shows similar plots for $NdFe_{1-y}Co_yAsO_{0.89}F_{0.11}$ samples,[19] for comparison.

Fig. 5. (color online) The main panel shows the Hall coefficient of $NdFe_{1-y}Ru_yAsO_{0.89}F_{0.11}$ against $T$ for various $y$ values, and the smaller panel shows similar plots for $NdFe_{1-y}Co_yAsO_{0.89}F_{0.11}$ samples,[19] for comparison.

Fig. 6. The superconducting transition temperatures of $NdFe_{1-y}Ru_yAsO_{0.89}F_{0.11}$ and $NdFe_{1-y}Co_yAsO_{0.89}F_{0.11}$ determined using the resistivity data are shown against $y$. These values agree reasonably well with the data obtained from the data of the superconducting diamagnetic moment.

Fig. 1

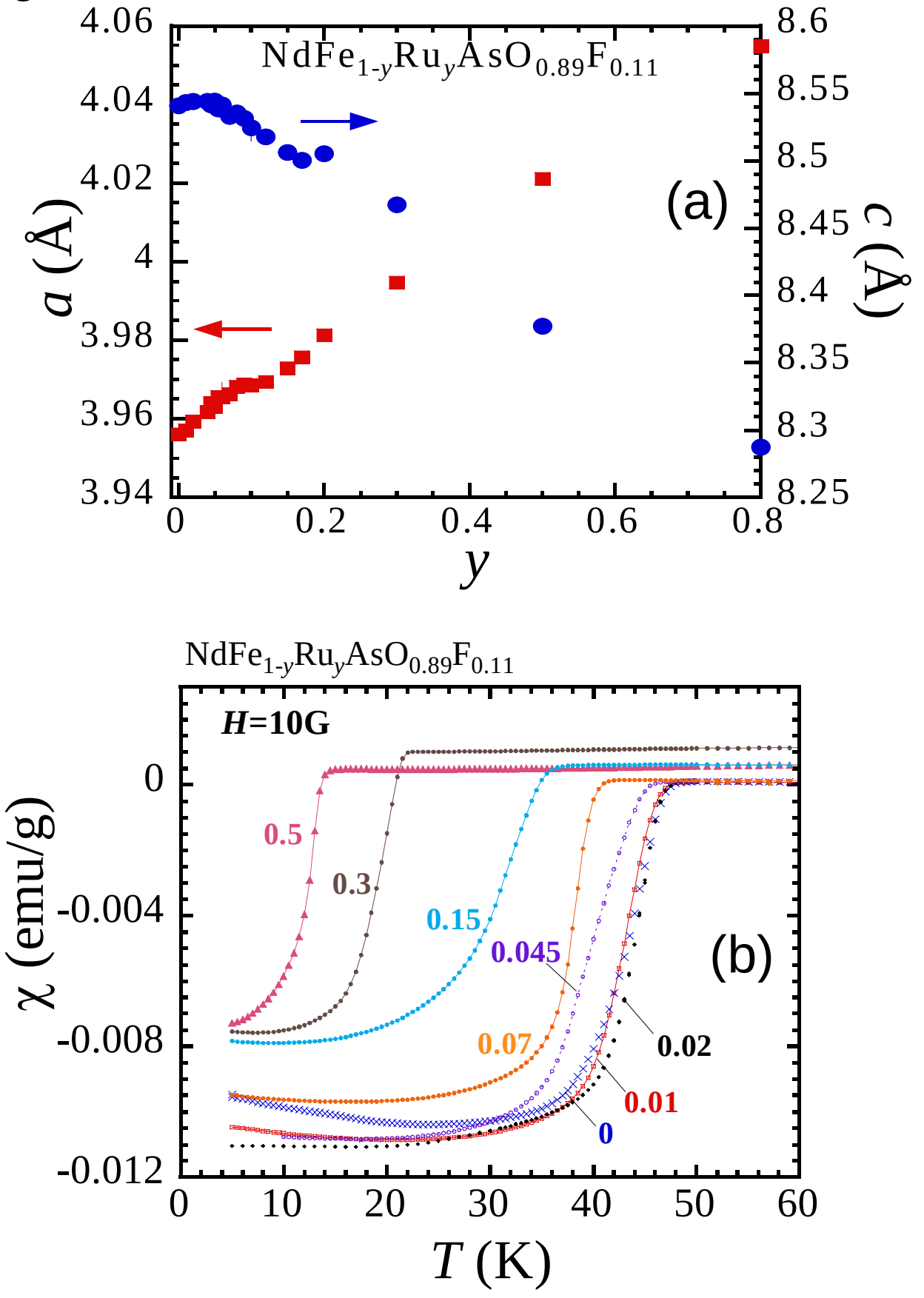

Fig. 3

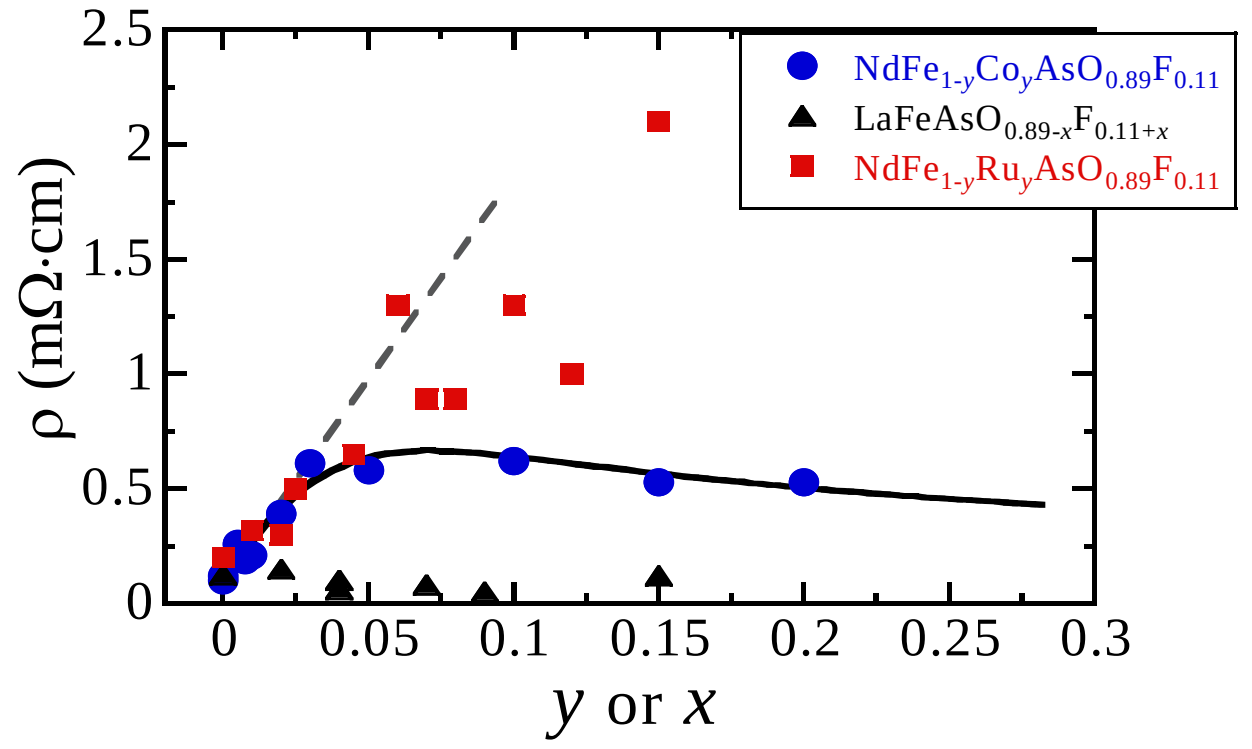

Fig. 2

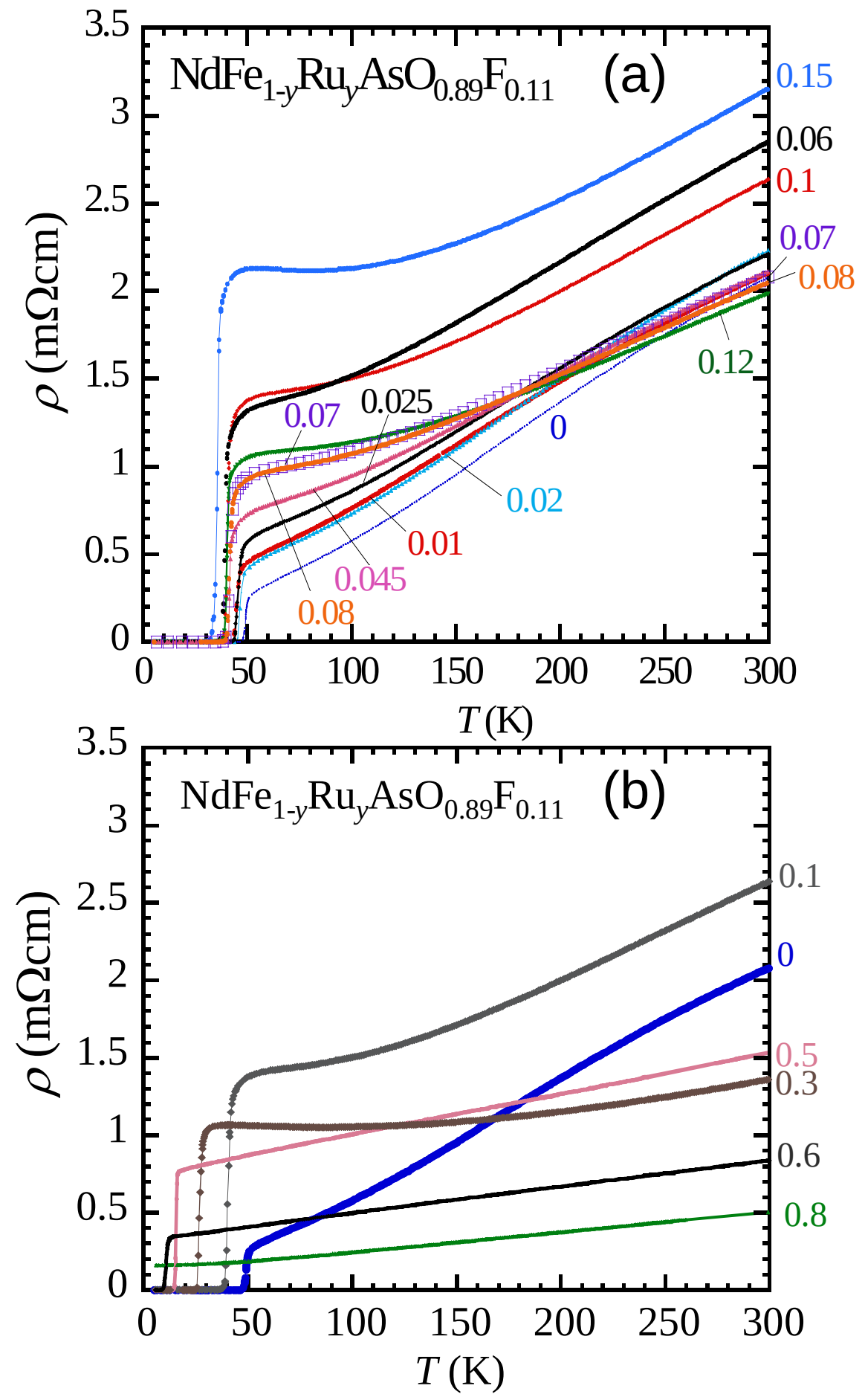

Fig. 4

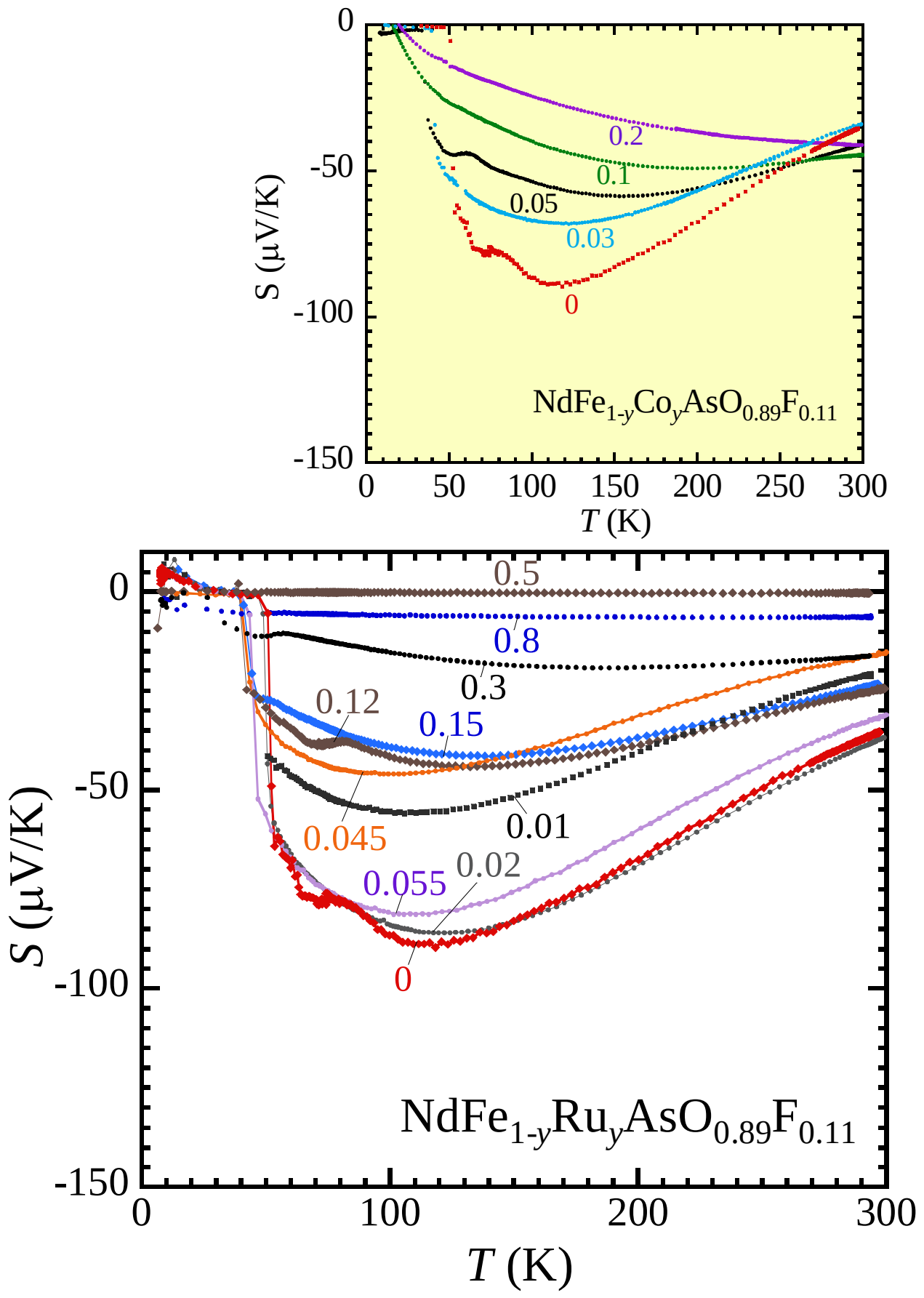

Fig. 5

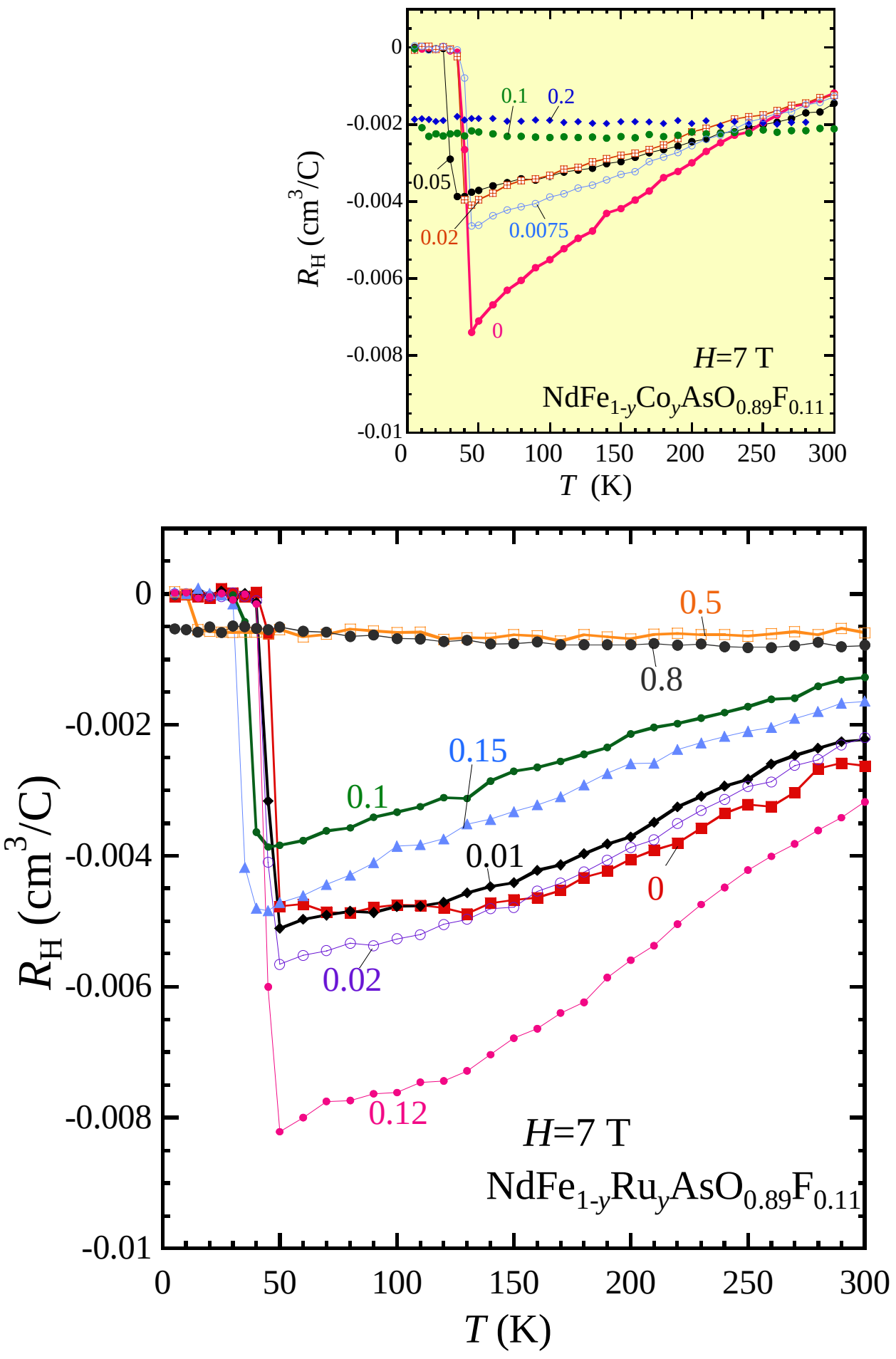

Fig. 6

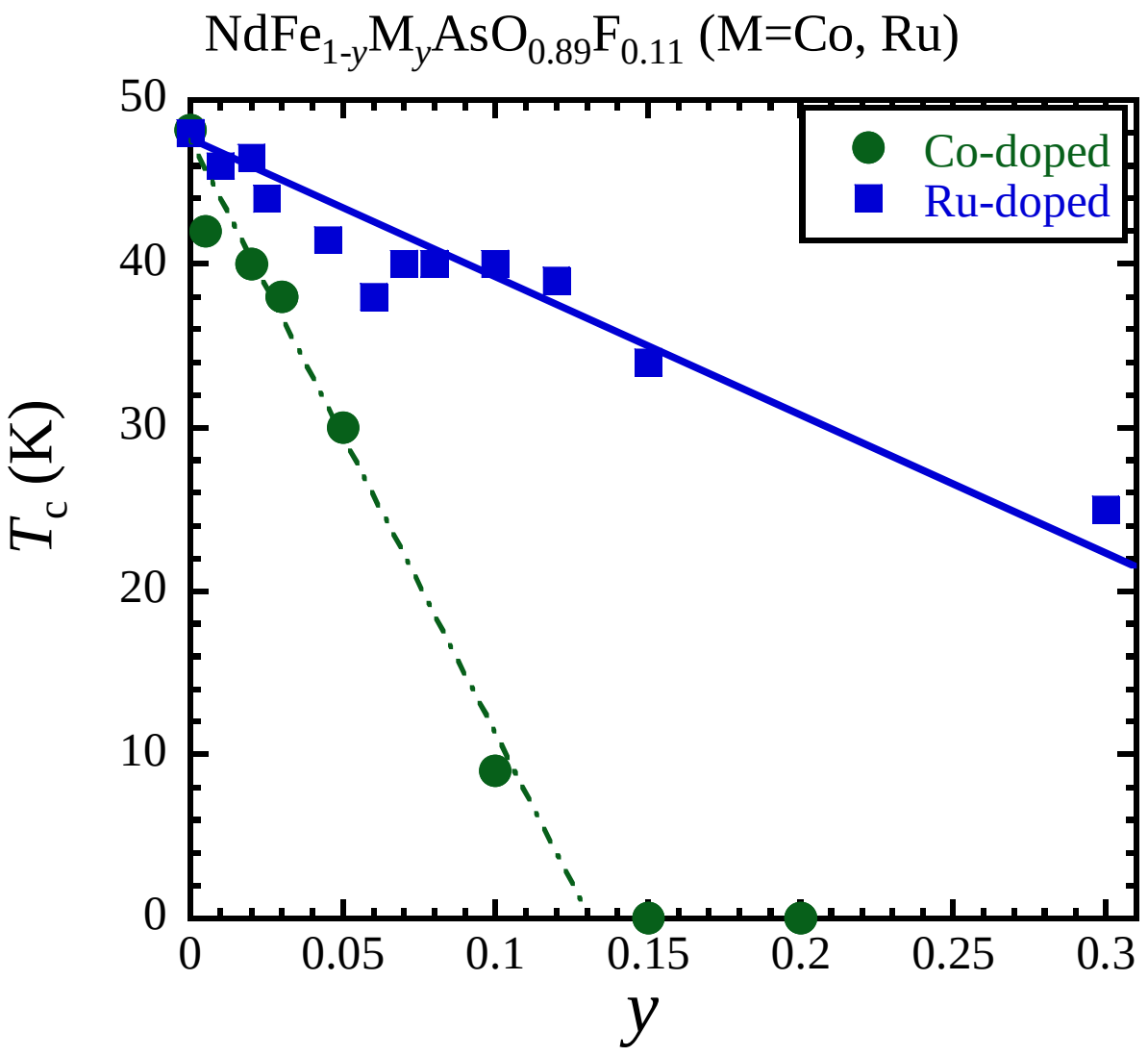

NdFe$_{1-y}$M$_y$AsO$_{0.89}$F$_{0.11}$ (M=Co, Ru)
Co-doped
Ru-doped
$T_c$ (K)
$y$